\documentclass[aps,prd,groupedaddress,showpacs,showkeys]{revtex4}
\newcommand{\eq}{\begin{eqnarray}}
\newcommand{\en}{\end{eqnarray}}

\def\rp{$R_p\hspace{-1em}/\ \ $}

\begin{document}

\title{Hadronic electric dipole moments in R-parity violating supersymmetry}
\author{
Amand \ Faessler$^1$, 
Thomas \ Gutsche$^1$, 
Sergey \ Kovalenko$^2$, 
Valery \ E. Lyubovitskij$^1$\footnote{On leave of absence
from Department of Physics, Tomsk State University,
634050 Tomsk, Russia}
\vspace*{1.2\baselineskip}}

\affiliation{$^1$ Institut f\"ur Theoretische Physik,
Universit\"at T\"ubingen,
\\ Auf der Morgenstelle 14, D-72076 T\"ubingen, Germany
\vspace*{1.2\baselineskip} \\
\hspace*{-1cm}$^2$ Departamento de F\'\i sica, Universidad
T\'ecnica Federico Santa Mar\'\i a, \\
Casilla 110-V, Valpara\'\i so, Chile
\vspace*{0.3\baselineskip}\\}

\date{\today}

\begin{abstract}
We calculate the electric dipole moments (EDM) of the neutral 
$^{199}$Hg atom, neutron and deuteron within a generic R-parity 
violating SUSY model (\rp SUSY) on the basis of a one-pion exchange 
model with CP-odd pion-nucleon interactions. We consider two types 
of the \rp SUSY contributions to the above hadronic EDMs:
via the quark chromoelectric dipole moments (CEDM) and 
CP-violating 4-quark interactions. We demonstrate that the former
contributes to all the three studied EDMs while the latter appears 
only in the nuclear EDMs via the CP-odd nuclear forces. 
We find that the \rp SUSY induced 4-quark interactions
arise at tree level through the sneutrino exchange and involve 
only $s$ and $b$ quarks. Therefore, their effect in hadronic EDMs 
is determined by the strange and bottom-quark sea of the nucleon. 
From the null experimental results on the hadronic EDMs 
we derive the limits on the imaginary parts of certain 
products Im$(\lambda^{\prime}\lambda^{\prime\ast})$ of the 
trilinear \rp-couplings and show that the currently best limits 
come from the $^{199}$Hg EDM experiments. We demonstrate that some 
of these limits are better than those existing in the literature. 
We argue that future storage ring experiments on the deuteron 
EDM are able to improve these limits by several orders of magnitude.
\end{abstract}



\maketitle


\section{Introduction} 
Over the years CP-violation (CPV) remains one of the central 
themes of particle physics and cosmology. By now the manifestations 
of CPV have been observed experimentally in the systems of 
K and B mesons. In the Standard Model (SM) there are two sources of 
CP-violation: the CKM phases and the QCD $\theta$-term. 
The former explains the observed CP-asymmetries in flavor changing 
K and B decays (for a recent review see, e.g.~\cite{CKM-KB-dec}) 
while the $\theta$-term, being flavor 
blind, is irrelevant for them but contributes to flavor neutral 
CP-odd observables such as electric dipole moments (EDM) of nucleons 
and nuclei (for review see, 
e.g.~\cite{CKM-KB-dec,Khriplovic-Lamoreaux,Khriplovich-PhRep}). 
Physics beyond the SM is introducing new complex parameters and, 
therefore, new sources of CP-violation in both flavor changing and 
neutral sectors. In supersymmetric (SUSY) models, these parameters come 
from the soft SUSY breaking sector and the superpotential $\mu$-term. 

In order to reveal the physics behind CP-violation one needs complementary 
information on various CP-odd observables. Among them the EDMs of leptons, 
nucleons and nuclei are attracting rising experimental and theoretical 
efforts as a sensitive probe of physics beyond the SM. During the last few 
years a significant progress has been achieved in experimental studies 
of various EDMs~\cite{neutron-exp}-\cite{deuteron-exp}. Presently there 
exist stringent upper bounds on the neutron 
EDM, $d_n$, \cite{neutron-exp_new} and the EDM, $d_{Hg}$, of 
the neutral $^{199}$Hg atom \cite{Hg-exp}: 
\eq\label{exp-dn} 
|d_n| &\leq& 3.0 \times 10^{-26} \ e\cdot {\rm cm} \,,\\
\label{exp-Hg} 
|d_{Hg}| &\leq& 2.1 \times 10^{-28} \ e\cdot {\rm cm} \ .
\en 
Recently it was also proposed to measure the deuteron EDM, $d_D$, 
in storage ring experiments~\cite{deuteron-exp} with deuteron ions 
instead of neutral atoms. The advantage of these experiments is the 
absence of Schiff screening, which introduces in significant uncertainties 
in the case of neutral atoms. This allows a direct probe of the $d_D$.  
In the near future it is hoped to obtain the experimental upper bound of 
\eq\label{exp-D} 
|d_D| \leq (1-3) \times 10^{-27} \ e\cdot {\rm cm}\ . 
\en 
The upper limits for the EDMs, derived from the above null experimental 
results, stringently constrain or even reject various models of 
CPV~\cite{Khriplovic-Lamoreaux}. 
For the case of the SUSY models with the superpartner masses around 
the electroweak scale $\sim$ 100 GeV $-$ 1 TeV, these limits imply the 
CPV SUSY phases to be very small. Various aspects of the calculation of 
the EDMs within the popular versions of SUSY models 
with~\cite{SUSY-EDM-1,Hisano} and 
without~\cite{BM86,SUSY-EDM-2,RPV-EDM,Herczeg-eN} R-parity conservation 
have been studied in the literature.

In the present paper we are studying the EDMs of the $^{199}$Hg atom and 
the deuteron as well as of the nucleon in the framework of a generic 
SUSY model without R-parity conservation (\rp SUSY) on the basis of 
chiral perturbation theory (ChPT). We consider CP-violation in the hadronic 
sector originating from the quark chromoelectric dipole moments (CEDMs) 
and 4-quark effective interactions which are induced by the complex phases 
of the trilinear \rp-couplings $\lambda^{\prime}$. 
From the experimental bounds of Eqs.~(\ref{exp-dn})-(\ref{exp-Hg})
we derive upper limits on the imaginary parts of the products 
$|{\rm Im} (\lambda^{\prime}_{i33} \lambda^{\prime\ast}_{i11})|$ and 
$|{\rm Im} (\lambda^{\prime}_{i22} \lambda^{\prime\ast}_{i11})|$ of 
the trilinear \rp-couplings and compare these limits to the existing 
ones. We also  discuss the prospects of the deuteron EDM experiments
from the view point of their ability to improve these limits.

\section{Hadronic EDMs: basic formalism}
\label{s2}

Here we briefly outline basic formulas for the calculation of the EDMs 
of the neutral $^{199}$Hg atom, the deuteron and nucleons. 
The $^{199}$Hg is a diamagnetic atom with a closed electron shell. 
Its EDM is dominated by the nuclear CP-violating effects characterized 
by the Schiff moment $S_{Hg}$, generating a T-odd electrostatic potential 
for atomic electrons. The $^{199}$Hg atomic EDM is given by~\cite{Hg} 
\eq\label{HgEDM} 
d_{Hg}= - 2.8\times 10^{-4} S_{Hg} \cdot {\rm fm}^{-2}\,. 
\en 
The deuteron EDM is a theoretically rather clean problem~\cite{Khriplovich} 
since the deuteron represents the simplest nucleus with a well understood 
dynamics. The corresponding EDM can be written as the sum of the three terms
\eq\label{dEDM-0}
d_D= d_p + d_n + d_D^{NN}\,,
\en 
where $d_n$, $d_p$ are the neutron and proton EDMs, respectively, and 
$d_D^{NN}$ is due to the CP-violating nuclear forces.

The EDMs of nucleons, $d_n$, $d_p$, the proton-neutron CP-odd interaction 
term, $d_D^{NN}$ and the Schiff moment $S_{Hg}$ can be evaluated on the 
basis of a one-pion exchange model with CP-odd pion-nucleon 
interactions~\cite{Khriplovich-PhRep,Khriplovich,DS:2005}.

The CP-odd pion-nucleon interactions are conventionally classified 
according to their isospin $(T)$ properties~\cite{CPV-pi-NN}:
\eq\label{CPV-1}
\Delta T = 0 \ {\rm transition}: \quad 
{\cal L}^{(0)}\,&=&  \,\bar{g}^{(0)} \, 
\overline{N} \, \vec{\pi} \, \vec{\tau} \, N \,,\\ 
\label{CPV-2}
|\Delta T| = 1 \ {\rm transition}: \quad 
{\cal L}^{(1)}\,&=& \, \bar{g}^{(1)} \, 
\overline{N} \, N \, \pi^0  \,, \\
\label{CPV-3}
|\Delta T| = 2 \ {\rm transition}: \quad 
{\cal L}^{(2)}\,&=&\, \bar{g}^{(2)} \, 
( \, \overline{N} \, \vec{\pi} \, \vec{\tau} \, N 
\, - \, 3 \, \overline{N} \, \tau^3 \, N \,\pi^0 \, ) \,, 
\en  
where the fields $N$ and $\vec{\pi}$ correspond to the 
doublet of nucleons and triplet of pions, respectively. 

For the calculation of the hadronic EDMs  
this Lagrangian is combined with 
the standard Lagrangian of the strong $\pi NN$ interaction
\eq\label{s}
{\cal L}_{\pi N} \, = \, g_{\pi N} \, \overline{N} \, 
i \, \gamma_{5} \, \vec{\pi} \, \vec{\tau} \, N \,.  
\en 
Here $g_{\pi N}$ is the CP-even $\pi NN$ coupling, which in 
the chiral limit satisfies the Goldberger-Treiman relation: 
\eq 
g_{\pi N} = g_A \frac{m_N}{F_\pi} = 12.9 \,,  
\en  
where $g_A = 1.267$ is the nucleon axial charge, 
$m_N = m_p = 938.27$ MeV 
is the nucleon mass and $F_\pi = 92.4$ MeV is the pion decay constant.  
In the adopted approach the nucleon EDMs are induced by the standard 
one-loop diagram with the charged $\pi$-meson which involves  
one CP-even and one CP-odd vertex corresponding to the terms in 
Eq.~(\ref{s}) and Eqs.~(\ref{CPV-1})-(\ref{CPV-3}), respectively.
In the leading order of the chiral expansion one finds \cite{pnEDM-Chiral}:
\eq\label{NEDM}
d_n = - d_p = 
\,\frac{e \, g_{\pi N} \, ( \bar g^{(0)} + \bar g^{(2)} )}{4\pi^2 m_N}\, 
\ln\frac{m_N}{M_{\pi}},
\en 
where $M_{\pi}= M_{\pi^+} = 139.57$ MeV is the $\pi$-meson mass. 
Thus, from Eqs.~(\ref{dEDM-0}) and (\ref{NEDM}) it follows that 
the nucleon contributions to the deuteron EDM cancel out in the leading 
order of the chiral expansion. This cancellation is a specific result 
of the SU(2) version of ChPT~\cite{EDMd-1} which does not hold in the  
SU(3) extension~\cite{Hisano}. In the present paper we do not consider the  
issues of the latter case. 

The pion exchange between two nucleons with one CP-even and one 
CP-odd vertex, corresponding to the terms in Eq.~(\ref{s}) and 
Eqs.~(\ref{CPV-1})-(\ref{CPV-3}), respectively, generates a P- and 
T-odd potential of proton-neutron forces. With this potential 
one can calculate the $d_D^{NN}$ term in Eq.~(\ref{dEDM-0}) as 
the expectation value of $e {\bf r}/2$, where ${\bf r}$ is the 
relative proton-neutron coordinate. In this way one obtains the 
following result~\cite{Khriplovich,Khriplovich-PhRep}:
\eq\label{dEDM-1}
d_D^{NN}\,=\,-\,{e \, g_{\pi N} \, \bar{g}^{(1)} \over 12 \pi M_{\pi}} 
\,{1+\xi\over (1+2\xi)^2} \,,
\en 
where $\xi = \sqrt{m_N E_B}/M_{\pi}$ and $E_B=2.23$ MeV is 
the deuteron binding energy. Numerically, one gets
\eq\label{dEDM-2}
d_D^{NN} = - 2.3\times 10^{-14}\,\bar{g}^{(1)}\;e\cdot  {\rm cm} \,.
\en 
Recently the Schiff moment $S_{Hg}$~\cite{DS:2005} has been calculated 
within a reliable nuclear structure model which takes full account of core 
polarization on the basis of the P- and T-odd one-pion exchange potential 
generated by the interactions in Eqs.~(\ref{CPV-1})-(\ref{s}). The result 
for a finite range interaction is 
\eq\label{Schiff-1}
S_{Hg}= - 0.055 \, g_{\pi N} \, (0.73 \times 10^{-2} 
\bar{g}^{(0)} + \bar{g}^{(1)} - 0.16 \bar{g}^{(2)}) 
\ e\cdot {\rm fm}^3.
\en 
Using the above equations one can derive from the experimental upper bounds 
in Eqs.~(\ref{exp-dn})-(\ref{exp-D}) the following constraints on the 
$\pi N$ CPV couplings:  
\eq\label{g-constr-Hg}
\mbox{present}\ \ d_{Hg}:\ 
|0.73 \times 10^{-2} 
\bar{g}^{(0)} + \bar{g}^{(1)} - 0.16 \bar{g}^{(2)})|  
&\leq& 10.6\times 10^{-12},\\ 
\label{g-constr-n}
\mbox{present}\  d_n\ : \hspace*{3.65cm} |\bar{g}^{(0)}+\bar{g}^{(2)}| 
&\leq& 2.3\times 10^{-12}, \\
\label{g-constr-D}
\mbox{future}\  d_D: \hspace*{4.65cm} |\bar{g}^{(1)}| 
&\leq& (0.43 - 1.3)\times 10^{-13}.
\en 
As evident, the future storage ring experiments with 
deuterons~\cite{deuteron-exp} are going to improve the limit on 
the CPV $\Delta T =1$ $\pi N$ coupling $\bar{g}^{(1)}$ by about 
two orders of magnitude.

In what follows, we calculate the \rp SUSY contributions to the quantities 
$S_{Hg}$, $d_n$, $d_p$ and $d_D^{NN}$ via the CP-odd $\pi NN$ 
interactions~(\ref{CPV-1})-(\ref{CPV-3}) and derive upper 
limits on  the \rp-couplings involved in these observables.

\section{CP-violating interactions in \rp SUSY}
\label{s3}

The effective CP-odd Lagrangian in terms of quark, gluon and photon 
fields up to operators of dimension~six, normalized at the hadronic scale 
$\sim$1 GeV, has the following standard form: 
\eq\label{CPV-q}
{\cal L}^{CPV} &=& \frac{\bar\theta}{16 \pi^{2}} {\rm tr}
\big(\widetilde{G}_{\mu\nu} G^{\mu\nu} \big) 
- \frac{i}{2} \sum_{i = u,d,s} d_i \ \bar q_i \, 
\sigma^{\mu \nu} \, \gamma_5 \, F_{\mu\nu} \, q_i 
- \frac{i}{2} \sum_{i = u,d,s} {\tilde d}_i\  \bar q_i \, 
\sigma^{\mu \nu} \, \gamma_5 
\, G_{\mu\nu}^a \, T^a \, q_i \,\nonumber\\
\hspace*{-2cm}
&-&  \frac{1}{6} \, C_W \,
f^{abc} \, G^a_{\mu\alpha}  \, G^{b\alpha}_{\nu}
\, G^c_{\rho\sigma} \, \varepsilon^{\mu\nu\rho\sigma} \,,
\en
where $G^a_{\mu\nu}$ is the gluon stress tensor,
$\widetilde{G}_{\mu\nu}=\frac{1}{2}
\epsilon_{\mu\nu\sigma\rho}{G}^{\sigma \rho}$ is its dual tensor,
and  $T^a$ and $f^{abc}$ are the $SU(3)$ generators and structure
constants, respectively. In this equation the first term
represents the SM QCD $\theta$-term, while the last three terms
are the non-renormalizable effective operators induced by physics
beyond the SM. The second and third terms are the dimension-five
electric and chromoelectric dipole quark operators, respectively,
and the last term is the dimension-six Weinberg operator. 
The light quark EDMs and CEDMs are denoted by $d_i$ and $\tilde{d}_i$,
respectively. 
In what follows we adopt the Peccei-Quinn mechanism, eliminating 
the $\bar\theta$-term as an independent source of CPV. 

We also consider the 4-quark 
CPV interactions of the form~\cite{4-fermion-1,4-fermion-2} 
\eq\label{4-q}
{\cal L}^{CPV}_{4q} = \sum_{i,j} \bigg\{
C^P_{ij}(\bar{q}_i q_i)(\bar{q}_j i \gamma_5 q_j) +
C^T_{ij}(\bar{q}_i \sigma_{\mu\nu} q_i)
(\bar{q}_j i \sigma^{\mu\nu} \gamma_5 q_j)\biggr\}\,,
\en 
where the sum runs over all the quark flavors $i,j=u,d,s,c,b,t$.
The operators of the above Lagrangians in Eqs.~(\ref{CPV-q})-(\ref{4-q}) 
can be induced by physics beyond the SM at loop or tree level after 
integrating out the heavy degrees of freedom.

We are studying the CPV effects in the hadronic sector induced by the 
trilinear interactions of the \rp SUSY models. The corresponding part 
of the $R_p$-violating superpotential reads: 
\eq\label{W_rp}
W_{R_p \hspace{-0.8em}/\;\:} = \lambda^{\prime}_{ijk}L_i Q_j D^c_k \,, 
\end{eqnarray} 
where the summation over the generation indices $i, j, k$ is understood, 
$L$, $Q$ and $D^c$ are the superfields of lepton-sleptons, 
quarks-squarks and $CP$-conjugated quarks-squarks, respectively,   
and $\lambda^{\prime}_{ijk}$ 
are the complex coupling constants violating lepton number conservation. 
Eq.~(\ref{W_rp}) results in the interactions
\eq\label{lambda}
{\cal L}_{\lambda^\prime} =
- \, \lambda _{ijk}^{\prime} \, ( \, 
\tilde{\nu}_{_{iL}}\bar{d}_{_k} P_{L} d_{_j} +
\tilde{d}_{_{jL}}\bar{d}_{_k} P_{L} \nu _{_i} +
\tilde{d}_{_{kR}}\bar{d}_{_j} P_{R} \nu^c_{_i}   
- \tilde{l}_{_{iL}}\bar{d}_{_k} P_{L} u_{_j}   
- \tilde{u}_{_{jL}}\bar{d}_{_k} P_{L} l_{_i}-
\tilde{d}_{_{kR}} \bar{u}_{_j} P_{R} l^c_{_i} \, ) \ +\ \mbox{H.c.}
\en 
with $P_{L,R} = (1 \mp \gamma_5)/2$. 

The interactions of the Lagrangian (\ref{lambda}) generate the terms in the 
effective CPV Lagrangians~(\ref{CPV-q}) and (\ref{4-q}) at certain orders in 
the $\lambda^{\prime}$-couplings.  
It is straightforward to derive the corresponding contribution to the 
4-quark contact terms (\ref{4-q}). It arises from a tree level contribution 
induced by the sneutrino, $\tilde\nu$, exchange given by 
\eq\label{4-q-RPV} 
{\cal L}^{CPV}_{4q} = [C^P_{sd} (\bar{s} s) 
                     + C^P_{bd} (\bar{b} b)] 
\ (\bar{d} i \gamma_5 d) \,
\en
with
\eq\label{C-P} 
C^P_{sd} = \sum\limits_{i} \frac{{\rm Im} (\lambda^{\prime}_{i22} 
\lambda^{\prime *}_{i11})}{2 m^2_{\tilde\nu(i)}}, \ \ \ \ \ 
C^P_{bd} = \sum\limits_{i} \frac{{\rm Im} (\lambda^{\prime}_{i33} 
\lambda^{\prime *}_{i11})}{2 m^2_{\tilde\nu(i)}}\,,
\en 
where $m_{\tilde\nu}$ is the sneutrino mass. Note, that the four-quark 
term involving only $d$-quarks is absent in (\ref{4-q-RPV}) due to 
${\rm Im} (\lambda^{\prime}_{i11} \lambda^{\prime *}_{i11}) \equiv 0$. 

The interactions of the Lagrangian~(\ref{lambda}) generate the quark EDMs, 
$d_q$, and CEDMs, $\tilde{d}_q$, starting from 
2-loops~\cite{SUSY-EDM-2,RPV-EDM} and the dominant \rp-contributions 
are of second order in the $\lambda^{\prime}$-couplings. 
It was shown in Ref. \cite{RPV-EDM} that the up-quark EDM and CEDM are 
suppressed by the light quark mass and mixing angles, which, therefore  
can be neglected. The quark EDMs are irrelevant for our study based on 
the pion-exchange model with the interaction the 
Lagrangian~(\ref{CPV-1})-(\ref{CPV-3}). We also do not consider the 
Weinberg term, which does not appear at the order of 
$O(\lambda^{\prime \, 2})$ unlike the quark CEDMs and 4-quark 
contact terms. In our analysis we use for the d-quark CEDMs the result 
of Ref.~\cite{RPV-EDM} 
\eq\label{CEDM-RPV}
\tilde{d}_d = Z^g \, \frac{\alpha_s}{32 \pi^3} \, \sum\limits_{i} \, 
\frac{m_b}{m^2_{\tilde\nu(i)}} \, 
F\left(\frac{m^2_b}{m^2_{\tilde\nu(i)}}\right) \, 
{\rm Im} (\lambda^{\prime}_{i33} \lambda^{\prime\ast}_{i11})\,,
\en 
with the loop function
\eq\label{F-loop}
F(x) \ = \ \int_{0}^{1} dy \ \frac{1-y(1-y)}{y(1-y)-x} \ 
\ln\frac{y(1-y)}{x} 
\ \rightarrow \   \frac{\pi^2}{3} +2 + \ln x + \ln^2x 
\quad\quad\quad \mbox{for}\ \ \  x \rightarrow 0\,. 
\en 
The expression~(\ref{CEDM-RPV}) corresponds to the d-quark CEDM 
at the hadronic scale $\sim$ 1 GeV. The renormalization group factor 
is evaluated to be $Z^g = 0.84$~\cite{RPV-EDM}. 

\section{\rp SUSY induced hadronic EDMs}
\label{s4}

We are studying the EDMs generated by pion-exchange with the CP-odd 
pion-nucleon interactions given by Eqs.~(\ref{CPV-1})-(\ref{CPV-3}). 
We assume that these interactions originate from the quark level effective 
Lagrangian~(\ref{CPV-q})-(\ref{4-q}). The corresponding relationships 
between the CPV $\pi$N-couplings, $\bar{g}^{(k)}$ 
of~(\ref{CPV-1})-(\ref{CPV-3}) and the parameters $\tilde{d}_q$, $C^P_{ij}$ 
of~(\ref{CPV-q})-(\ref{4-q}) can be conventionally extracted from the 
matching condition 
\eq\label{Matching} 
\langle \pi^0 N|{\cal L}^{CPV} + {\cal L}^{CPV}_{4q}|N \rangle = 
\langle \pi^0 N| {\cal L}^{(0)} + {\cal L}^{(1)} 
+ {\cal L}^{(2)}|N \rangle\,. 
\end{eqnarray}
In Ref.~\cite{Hisano} the contributions of the quark CEDMs 
to the couplings $\bar{g}^{(k)}$ were determined on the basis of the 
chiral Lagrangian and of the phenomenological evaluation of the nucleon 
matrix elements in the l.h.s. of Eq.~(\ref{Matching}). 
There appear non-zero contributions only to the 
$\bar{g}^{(0,1)}_{_{CEDM}}$ couplings 
\eq\label{g0-CEDM}
\bar{g}^{(0)}_{_{CEDM}} &=& - 1.02  \times 
\left(\frac{\tilde{d}_d}{{\rm GeV}^{-1}}\right) 
= - 6.3 \times 10^{-7} \, \sum_{i} \,  
{\rm Im} (\lambda^{\prime}_{i33} \lambda^{\prime\ast}_{i11})\,, \\
\label{g1-CEDM} 
\ \bar{g}^{(1)}_{_{CEDM}} &=& 9.2 \times 
\left(\frac{\tilde{d}_d}{{\rm GeV}^{-1}}\right)
= 5.7 \times 10^{-6} \, \sum_{i} \,  
{\rm Im} (\lambda^{\prime}_{i33} \lambda^{\prime\ast}_{i11})\,.
\en 
Here we disregarded the contribution 
of $\tilde{d}_u$ as it is strongly suppressed in \rp SUSY~\cite{RPV-EDM}. 
 
Due to its isospin structure, the effective 4-quark interaction in 
Eq.~(\ref{4-q-RPV}) contributes only to the $|\Delta T| = 1$  
coupling $\bar{g}^{(1)}$ in Eq.~(\ref{CPV-2}). 
We derive the corresponding contribution using Eq.~(\ref{Matching}). 
In order to evaluate the matrix element in the l.h.s. of~(\ref{Matching}) 
we apply the relation of partial conservation of the axial current (PCAC) 
\eq\label{PCAC} 
\langle \pi^0 |\bar{d} i \gamma_5 d|0 \rangle = 
- F_{\pi} \frac{M^2_{\pi}}{2 m_d}\,, 
\en 
where $m_d = 9$ MeV is the $d$-quark current mass.
We also need to know the nucleon matrix elements of the scalar currents
\eq\label{NNs} 
\langle N |\bar{s} s|N \rangle = G^{(s)}_S, \ \ \ \  
\langle N |\bar{b} b|N \rangle = G^{(b)}_S\,.
\en 
The values of $G^{(s)}_S$, $G^{(b)}_S$ are subject to significant 
uncertainties. In our analysis we use the estimates from 
Refs.~\cite{Gsb1,Gsb2} 
\eq\label{ffn} 
G^{(s)}_S = (0.64-3.9), \ \ \ \  G^{(b)}_S = 9\times 10^{-3}\,. 
\en 
For the value of $G^{(s)}_S$ we indicate the interval of possible values 
according to Ref.~\cite{Gsb1}. For $G^{(b)}_S$ we only need an order of 
magnitude estimate which will be sufficient to conclude that it is associated 
with the subdominant term not essential for our analysis.

From the matching condition (\ref{Matching}) and using Eqs.~(\ref{4-q}), 
(\ref{C-P}) and (\ref{PCAC})-(\ref{ffn}) we find for 
the contribution of the 4-quark interactions 
\eq\label{g-1-2} 
\bar{g}^{(1)}_{4q} &=& 
- F_{\pi} \frac{m_{\pi}^2}{2m_d}\left(C^P_{sd} G^{(s)}_S + 
C^P_{bd} G^{(b)}_S\right) = \nonumber\\
&=& - (3.6 - 21.7)\times 10^{-9} \, \sum\limits_{i} \, 
{\rm Im} (\lambda^{\prime}_{i22} 
\lambda^{\prime\ast}_{i11}) - 5 \times 10^{-11} 
\, \sum\limits_{i} \, 
{\rm Im} (\lambda^{\prime}_{i33} \lambda^{\prime\ast}_{i11})\,.  
\en 
Eqs.~(\ref{g0-CEDM}), (\ref{g1-CEDM}) and (\ref{g-1-2}) show that 
the CPV $\pi N$ couplings  $\bar{g}^{(0)}=\bar{g}^{(0)}_{_{CEDM}}$ and 
$\bar{g}^{(1)}=\bar{g}^{(1)}_{_{CEDM}}+\bar{g}^{(1)}_{4q}$  
receive the contribution from 
${\rm Im} (\lambda^{\prime}_{i33} \lambda^{\prime\ast}_{i11})$ dominated 
by the $d$-quark CEDM while the contribution from 
${\rm Im} (\lambda^{\prime}_{i22} \lambda^{\prime\ast}_{i11})$
appears solely via the 4-quark CPV interactions~(\ref{4-q-RPV}).

\begin{table}[b]
\label{Table-1}
\caption{
Upper limits on the imaginary parts of the products of the trilinear 
\rp-couplings derived from the experimental bounds on the EDMs of 
neutron \protect\cite{neutron-exp_new},  
$^{199}$Hg neutral atom \protect\cite{Hg-exp} 
and deuteron \protect\cite{deuteron-exp}. 
The existing constraints from other experiments on the absolute values 
of the corresponding products of \rp-coupling are taken from 
Ref.~\protect\cite{RPV-rev}. The limits are evaluated with all the 
superpartner masses equal to 300 GeV.} 

\def\arraystretch{1.}
\begin{center}
\begin{tabular}{|c|c|c|c|c|}
\hline
Experiment& $d_n$ \protect\cite{neutron-exp_new} 
          & $d_{Hg}$ \protect\cite{Hg-exp} 
          & $d_D$ \protect\cite{deuteron-exp}
          & Existing limits \protect\cite{RPV-rev} \\
\hline
$\sum\limits_{i} \, |{\rm Im} (\lambda^{\prime}_{i33} 
\lambda^{\prime\ast}_{i11})|$& 
$3.6\times 10^{-6}$ & $1.9\times 10^{-6}$&
$(0.8 - 3.2)\times 10^{-8}$ & 
$|\lambda^{\prime}_{133} \lambda^{\prime}_{111}|\leq 4.5\times 10^{-5}$ 
\\
&&&& 
$|\lambda^{\prime}_{233} \lambda^{\prime}_{211}|\leq 5.4\times 10^{-3}$\\ 
[2mm] 
&&&& $|\lambda^{\prime}_{333} \lambda^{\prime}_{311}|\leq 1.3\times 10^{-3}$\\
\hline
$\sum\limits_{i} \, 
|{\rm Im} (\lambda^{\prime}_{i22} \lambda^{\prime\ast}_{i11})|$ & 
- &$(0.3-1.9)\times 10^{-3}$ & 
$(2-36.1)\times 10^{-6}$& 
$|\lambda^{\prime}_{122} \lambda^{\prime}_{111}|\leq 4.5\times 10^{-5}$ 
\\
&&&& $|\lambda^{\prime}_{222} \lambda^{\prime}_{211}|\leq 1.3\times 10^{-3}$\\
[2mm]
&&&& $|\lambda^{\prime}_{322} \lambda^{\prime}_{311}|\leq 1.3\times 10^{-3}$\\
\hline
\end{tabular}
\end{center}
\end{table}

Now we are ready to derive the constraints on the trilinear \rp-couplings 
from the experimental bounds of Eqs. (\ref{g-constr-Hg})-(\ref{g-constr-D}). 
Using the expressions (\ref{g0-CEDM}),(\ref{g1-CEDM}) and (\ref{g-1-2}) 
we obtain from the above experimental bounds the upper limits on 
$|{\rm Im} (\lambda^{\prime}_{i33} \lambda^{\prime\ast}_{i11})|$ and 
$|{\rm Im} (\lambda^{\prime}_{i22} \lambda^{\prime\ast}_{i11})|$. 
These limits, extracted from  the EDMs of the $^{199}$Hg atom, neutron
and deuteron, are given in Table I together with the existing limits
on $|\lambda^{\prime}_{i33} \lambda^{\prime\ast}_{i11}|$ and
$|\lambda^{\prime}_{i22} \lambda^{\prime\ast}_{i11}|$
\cite{RPV-rev}.
The limits from the EDMs of $^{199}$Hg and the deuteron are shown for the
uncertainty interval of the scalar nucleon form factor
introduced in Eq. (\ref{ffn}).
It is seen that the presently most stringent limits on $|{\rm Im}
(\lambda^{\prime}_{ikk} \lambda^{\prime\ast}_{i11})|$
come from the $^{199}$Hg atom EDM (\ref{exp-Hg}). The forthcoming 
experiments on the deuteron EDM
(\ref{exp-D}) are going to improve these limits by about one to three
orders of magnitude. 
Note, that we obtained about 1-order of magnitude improvement for the limit 
$|{\rm Im} (\lambda^{\prime}_{i33} 
\lambda^{\prime\ast}_{i11})|\leq 1.2\times 10^{-5}$ previously derived
in Ref. \cite{RPV-EDM} from the neutron EDM constraint~(\ref{exp-dn}) 
on the basis of the SU(6) quark model. 
The existing limits on the absolute values of the corresponding products 
of the $\lambda^{\prime}$-couplings depreciate the present EDM limits on 
$|{\rm Im} (\lambda^{\prime}_{i22} \lambda^{\prime\ast}_{i11})|$, being more 
stringent, while they do not exclude the values of 
$|{\rm Im} (\lambda^{\prime}_{i33} \lambda^{\prime\ast}_{i11})|$ within the 
limits derived from EDMs. We also observe that the future deuteron EDM 
experiments are expected to decrease the upper limits on 
$|{\rm Im} (\lambda^{\prime}_{i22} \lambda^{\prime\ast}_{i11})|$ below the 
values excluded by the existing limits on the absolute values of these 
products. 

\section{Summary and Conclusions}
\label{summary}

We have studied the contributions of the trilinear \rp-couplings 
to the $^{199}$Hg atom, neutron and deuteron EDMs within ChPT, 
applying the pion-exchange model 
of CPV nuclear forces. We have analyzed the \rp-contributions via 
the d-quark CEDM and CPV 4-quark interactions.
We have shown that the \rp SUSY induced 4-quark interactions 
contributes only to the nuclear EDMs such as $^{199}$Hg and deuteron EDMs via 
the CPV nuclear forces and do not contribute to the neutron EDM. 
We have also found that
these two type of mechanisms give rise to a dependence of the hadronic EDMs 
proportional to different $\lambda^{\prime}$-couplings.
Therefore, taking into account both mechanism allows one 
to obtain a complimentary information on the imaginary parts of 
the products of the $\lambda^{\prime}$-couplings. The corresponding 
upper limits from the null experimental results on measurements of 
the above mentioned hadronic EDMs are given in Table 1.
We have demonstrated that the present limits from the $^{199}$Hg EDM 
experiments are by a factor $\sim$6 more stringent than those from the 
experiments on the neutron EDM and that the planned storage ring experiments 
with the deuterium ions would be able to significantly improve these limits.

\section{Acknowledgments}

This work was supported in part by the FONDECYT project 1030244,
by the DFG under the contracts FA67/25-3 and GRK683.
This research is also part of the EU Integrated Infrastructure
Initiative Hadronphysics project under the contract number
RII3-CT-2004-506078 and President grant of Russia "Scientific
Schools"  No 5103.2006.2.

\end{document}